\documentclass[a4paper,twocolumn,english,aps,showpacs]{revtex4}
\usepackage[T1]{fontenc}
\usepackage[latin9]{inputenc}
\usepackage{float}
\usepackage{units}
\usepackage{amsmath}
\usepackage{amssymb}
\usepackage{graphicx}
\usepackage{esint}

\makeatletter

\@ifundefined{textcolor}{}
{%
 \definecolor{BLACK}{gray}{0}
 \definecolor{WHITE}{gray}{1}
 \definecolor{RED}{rgb}{1,0,0}
 \definecolor{GREEN}{rgb}{0,1,0}
 \definecolor{BLUE}{rgb}{0,0,1}
 \definecolor{CYAN}{cmyk}{1,0,0,0}
 \definecolor{MAGENTA}{cmyk}{0,1,0,0}
 \definecolor{YELLOW}{cmyk}{0,0,1,0}
}

\usepackage{babel}

\usepackage{babel}

\makeatother

\usepackage{babel}
\begin{document}

\title{Semiclassical Coherent States propagator }

\author{Alejandro M.F Rivas}

\thanks{member of the CONICET}

\date{\today {}}
\begin{abstract}
In this work, we derived a semiclassical approximation for the matrix
elements of a quantum propagator in coherent states (CS) basis that
avoids complex trajectories, it only involves real ones. For that
propose, we used the, symplectically invariant, semiclassical Weyl
propagator obtained by performing a stationary phase approximation
(SPA) for the path integral in the Weyl representation. After what,
for the transformation to CS representation SPA is avoided, instead
a quadratic expansion of the complex exponent is used. This procedure
also allows to express the semiclassical CS propagator uniquely in
terms of the classical evolution of the initial point, without the
need of any root search typical of Van Vleck Gutzwiller based propagators.
For the case of chaotic Hamiltonian systems, the explicit time dependence
of the CS propagator has been obtained. The comparison with a \textquotedbl{}realistic\textquotedbl{}
chaotic system that derives from a quadratic Hamiltonian, the cat
map, reveals that the expression here derived is exact up to quadratic
Hamiltonian systems. 
\end{abstract}

\affiliation{Departamento de Física, Comisión Nacional de Energía Atómica. Av.
del Libertador 8250, 1429 Buenos Aires, Argentina. \\
 }


\maketitle

\section{Introduction}

\label{sec1}

Path integrals appear as a useful calculation tool for many quantum
and statistical mechanical problems \cite{path}, while CS are widely
known to represent quantum states with the most classical resemblance.
In the case of an harmonic oscillator, they obey the classical equations
of motion and are minimal uncertain states. Also, the CS form an overcomplete
basis which is a necessary ingredient for the construction of a path
integral \cite{coherent}. It implies the existence of several forms
of path integrals, all quantum mechanically equivalent, but each leading
to a different semiclassical limit. The formulation of path integrals
applied to CS has become widely used in many areas despite the very
shaky mathematical background \cite{klauderarx}.

For the propagator in a mixed or coordinates representation, the so-called
Herman\textendash{}Kluk (HK) formula and its generalizations \cite{herman,miller,pollak},
semiclassicaly derived in \cite{kay,kay2}, are routinely used nowadays
and expresses the propagator as an integral over the overcomplete
basis of CS. While for the case of the propagator in CS representation,
a complete semiclassical derivation was performed by Baranger \textit{et
al} \cite{Baranger}, also in \cite{aguiar} a Weyl ordering treatment
has been performed. Although mathematically correct, both constructions
\cite{Baranger} and \cite{aguiar}  involve an analytic continuation
to complex trajectories, while the classical system only involves
real canonical variables. As was recently pointed out, the CS path
integral breaks down in certain cases \cite{pathPRL}. When the Hamiltonian
involves terms that are non linear in generators, neither the action
suggested by Weyl ordering nor the one constructed by normal ordering
gives correct results. In order to understand the quantum classical
limit it is imperious to have a correct semiclassical expression of
the quantum propagator in the most classical states, that is, in CS.

In this work, we derive an accurate semiclassical expression for the
CS propagator that avoids complex trajectories, it only involves real
ones. While the, symplectically invariant, semiclassical Weyl propagator
is obtained by performing a SPA from the path integral in the Weyl
representation, for the transformation to CS representation SPA is
avoided, instead a quadratic expansion of the complex exponent is
used. This procedure also allows to express the semiclassical CS propagator
uniquely in terms of objects obtained directly by the classical evolution
of the flux through the initial point, without the need of any further
trajectory search that are typical procedures of Van Vleck Gutzwiller
(VVG) based propagators \cite{gutz2,VVleck}. Also, for the case of
chaotic Hamiltonian systems, the explicit time dependence of the CS
propagator has been obtained only in terms of the action of the orbit
through the initial point, the Lyapunov exponent and the stable and
unstable directions. The comparison with a \textquotedbl{}realistic\textquotedbl{}
chaotic system, the cat map, reveals that the expression (\ref{eq:XUXSCLIN})
here derived is exact up to quadratic Hamiltonian systems. While,
in the CS representation, a common SPA does not give accurate results
with real trajectories.

This paper is organized as follows: In section \ref{sec2} we introduce
Weyl Wigner representation of the quantum propagator and its close
connection with the center generating function of the canonical transformations
of classical mechanics \cite{ozrep} that will be used all along this
work. Then we study the coherent states propagator obtained from a
path integration in the Weyl Wigner representation by means of the
SPA. We will see in section \ref{sec4} that the expressions obtained
in this way failed to give accurate results.

For section \ref{sec3} instead of a simple SPA we expand the center
action up to quadratic terms and there after we perform the phase
space integration from the Weyl propagator to the CS propagator. In
this way, we obtain a general semiclassical expression of the matrix
elements of the propagator in the CS basis that only involves real
trajectories. We show how the quadratic expansion of the center generating
function allowed to express the semiclassical CS propagator uniquely
in terms of objects obtained directly by the classical evolution of
the flux through the initial point, without the need of any further
search of trajectories. In this section we also perform a comparison
of the here derived expression with the well known Initial Value Representation
(IVR) methods that includes the HK propagator and Heller`s Thawed
Gaussian Approximation (TGA).

In Section \ref{sec4}, the study of continuous Hamiltonian systems
is projected to maps on Poincaré's sections. The explicit time dependence
of the CS propagator is obtained only in terms of the action of the
orbit through the initial point, the Lyapunov exponent and the stable
and unstable directions. Then we study the particular case of the
cat map where not only the semiclassical theory is exact but also
the linear approximation is valid throughout the torus. After the
semi classical expressions here deduced are adapted for a torus phase
space, we then see that the expression obtained in section \ref{sec3}
gives results that coincides exactly with the numerically computed
matrix elements for the cat maps. However the expressions deduced
in section \ref{sec2} from a common SPA do not give accurate results
with real trajectories. Finally, we give our conclusions in Section
\ref{sec5}. The appendix is devoted to the Weyl Wigner representation
of quantum mechanics in terms of reflections in phase space.


\section{Coherent States propagator and stationary phase }

\label{sec2}

The coherent state $\left|X\right\rangle $, centered on the point
$X=\left(P,Q\right)$ in phase space, is obtained by translating the
ground state of the harmonic oscillator, its position representation
is
\begin{equation}
\left\langle q\!\right.\left|X\!\right.\rangle=\!\left(\frac{m\omega}{\pi\hbar}\right)^{\frac{1}{4}}\!\exp\!\left[\!-\frac{\omega}{2\hbar}(q-Q)^{2}\!+\! i\frac{P}{\hbar}\left(q\!-\frac{Q}{2}\right)\!\right].\label{eq:qX}
\end{equation}
 Without loss of generality, unit frequency $\left(\omega=1\right)$
and mass $\left(m=1\right)$ are chosen for the harmonic oscillator.
The overlap of two CS is then 
\begin{equation}
\langle X\left|X^{\prime}\rangle\right.=\exp\left[-\frac{\left(X-X^{\prime}\right)^{2}}{4\hbar}-\frac{i}{2\hbar}X\wedge X^{\prime}\right],\label{cohcoh}
\end{equation}
 with the wedge product 
\[
X\wedge X^{\prime}=PQ^{\prime}-QP^{\prime}=\left({\mathcal{J}}X\right).X^{\prime}.
\]
 The second equation also defines the symplectic matrix ${\mathcal{J}}$,
that is 
\begin{equation}
{\mathcal{J}}=\left[\begin{array}{cc}
0 & -1\\
1 & 0
\end{array}\right].
\end{equation}
 In what follows, we will use the centers and chords formulation developed
by Ozorio de Almeida \cite{ozrep} for both classical and quantum
mechanics.

Let us write the quantum propagator $\widehat{U}^{t}$ in terms of
its, symplectically invariant, center or Weyl Wigner symbol $U^{t}(x)$
\cite{ozrep}, 
\begin{equation}
\widehat{U}^{t}\!=\!\frac{1}{\left(\pi\hbar\right)^{l}}\!\int\! dx\widehat{R}_{x}U^{t}(x)\!\quad\!\textrm{and }\!\quad\! U^{t}(x)\!=\! tr\!\left[\widehat{R}_{x}\widehat{U}^{t}\right]\!,\label{eq:SWF-1}
\end{equation}
 where$\int\! dx$ is an integral over the whole phase space of $l$
degrees of freedom, while $\widehat{R}_{x}$ denotes the set of reflection
operators thought points $x=\left(p,q\right)$ in phase space \cite{ozrep}
(see Appendix). Hence, the matrix elements of the unitary propagator
$\widehat{U}^{t}$, that governs the quantum evolution of an $l$
degrees of freedom system, in the CS basis, are 
\begin{equation}
\langle X_{1}|\widehat{U}^{t}|X_{2}\rangle=\left(\frac{1}{\pi\hbar}\right)^{l}\int\langle X_{1}|U^{t}(x)\widehat{R}_{x}|X_{2}\rangle dx.\label{eq:XURX}
\end{equation}
 The action of a reflection operator $\widehat{R}_{x}$ on a coherent
state $\left|X\right\rangle $ is the $x$ reflected coherent state
(see Appendix) 
\begin{equation}
\widehat{R}_{x}\left|X\right\rangle =e^{\frac{i}{\hbar}X\wedge x}\left|2x-X\right\rangle .\label{reflexcoh}
\end{equation}
 Inserting (\ref{reflexcoh}) and (\ref{cohcoh}) in expression (\ref{eq:XURX})
the propagator in the CS representation is obtained from the Weyl
propagator as 
\begin{equation}
\langle X_{1}|\widehat{U}^{t}|X_{2}\rangle\!=\! e^{\!\frac{iX_{1}\!\wedge\! X_{2}}{2\hbar}}\!\!\int\!\!\frac{dx}{\left(\pi\hbar\right)^{l}}e^{\left[\frac{i}{\hbar}x\wedge\xi_{0}-\frac{\left(X-x\right)^{2}}{\hbar}\right]}U^{t}(x),\label{eq:XUgaussX}
\end{equation}
 with $\xi_{0}\equiv\left(X_{1}-X_{2}\right)$ the chord joining the
points $X_{1}$ and $X_{2}$, while $X=\nicefrac{1}{2}\left(X_{1}+X_{2}\right)$
denotes their mid point. The CS representation is overcomplete. Indeed,
just the diagonal elements $\langle X|\widehat{U}^{t}|X\rangle$ known
as the Husimi representation, are enough for a complete representation
of quantum mechanics. As equation (\ref{eq:XUgaussX}) shows, the
Husimi representation is a Gaussian smoothing of the Weyl-Wigner's.

For a Hamiltonian system, the propagator is $\widehat{U}^{t}=\exp[-i\frac{t}{\hbar}\widehat{H}]$
with $\widehat{H}$ the quantum Hamiltonian operator of the system.
As has been shown in \cite{ozrep}, for sufficiently small times,
the Weyl symbol of the propagator is 
\begin{equation}
U^{t}(x)=\!|\det[1\!-\!(\mathfrak{J}\frac{t}{2}{\cal H})^{2}]|^{\frac{1}{4}}\!\exp\!\left[\!-\frac{it}{\hbar}H(x)\right]\!+\! O(\left(t\hbar\right)^{2}),\label{eq:UxSC2}
\end{equation}
 where $H(x)=tr\left[\widehat{R}_{x}\widehat{H}\right]$ is the Weyl
symbol of the Hamiltonian operator $\widehat{H}$, while ${\cal H}$
is the corresponding Hessian matrix evaluated at the point $x$. In
analogy to the classical theory, $H(x)$ is, within $O(\hbar)$, the
classical Hamiltonian $H_{c}(x)$. Also, if $H(x)$ is quadratic (\ref{eq:UxSC2})
represents a unitary operator.

For longer times $t$, the composing of an even number $k$ of unitary
operators for times $\varepsilon=t/k$ is performed. In the Weyl representation,
such a composition results in \cite{ozrep}
\begin{equation}
U^{t}(x)=\int\frac{dx_{i}^{k}}{(\pi\hbar)^{kl}}\prod_{j=1}^{k}U^{\frac{t}{k}}(x_{j})e^{\frac{i}{\hbar}\Delta_{k+1}(x,x_{1},\cdots,x_{k})},\label{eq:UNx}
\end{equation}
 with $dx_{i}^{k}=dx_{k}\cdots dx_{1}$ and $\Delta_{k+1}(x,x_{1},\cdots,x_{k})$
denotes the symplectic area of the polygon with endpoints centered
on $x$ and whose $j'$th side is centered on $x_{j}$. The formula
(\ref{eq:UNx}) is exact, while in the limit $k\rightarrow\infty$,
the expression (\ref{eq:UxSC2}) can be inserted for $U_{\frac{t}{k}}(x)$.
In which limit, the amplitude vanishes, yielding to the \textit{path
integral}
\begin{equation}
\! U^{t}(x)\!=\!\lim_{k\rightarrow\infty}\!\!\int\!\!\frac{dx_{i}^{k}}{(\pi\hbar)^{kl}}e^{\frac{i}{\hbar}\left[\!\Delta_{k+1}\!(\! x,x_{1},\!\cdots\!,x_{k}\!)-\frac{t}{k}\!\sum_{j=1}^{k}\! H(x_{j})\right]}\!.\label{eq:UxPath}
\end{equation}
 Since (\ref{eq:UxPath}) is an ordinary multiple integral, we need
not worry about the definition of the ``path space'' \cite{ozrep}.

Analogously, for classical mechanics, the center generating function
of the canonical transformation resulting from composing of an even
number $k$ of canonical transformations for time $\varepsilon=t/k$
in the limit where $k\rightarrow\infty$, is \cite{ozrep} 
\begin{equation}
S^{t}(x)=\!\lim_{k\rightarrow\infty}\!-\frac{t}{k}\!\sum_{j=1}^{k}\! H_{c}(x_{j})\!+\!\Delta_{k+1}\!(\! x,x_{1},\!\cdots\!,x_{k}\!).\label{eq:CGF}
\end{equation}
 The polygon $\Delta_{k+1}$ has one large side $\xi$ passing through
the center $x$ and $k$ small chords tangents to the orbit as $k\rightarrow\infty$
such that $\partial S^{t}/\partial x_{j}=0$. Hence, the \textit{center
variational principle} is obtained: The \textit{center action} 
\begin{equation}
S_{\gamma}^{t}(x)=\oint_{x}p\cdot dq-\int_{\gamma}H_{c}(x)dt
\end{equation}
 is stationary along the classical trajectory $\gamma$. The paths
to be compared always have their endpoints centered on the point $x$.
The second integral is evaluated along this path, whereas the first
integral is closed off by the chord centered on $x$. Then $S_{\gamma}^{t}(x)$
is the classical center generating function of the classical trajectory
$\gamma$, from which the chord $\xi$ joining the initial and final
point of the trajectory is obtained 
\begin{equation}
\xi=-{\cal J}\frac{\partial S_{\gamma}^{t}(x)}{\partial x}.\label{cuerda}
\end{equation}

The semiclassical approximation consist in evaluating the path integral
in the SPA. The result is the leading term in a series of increasing
powers of $\hbar$ . For the Weyl propagator this was obtained in
\cite{ozrep} by performing a SPA in (\ref{eq:UxPath}). The phase
of the integral in (\ref{eq:UxPath}) coincides with the center action
$S_{\gamma}^{t}(x)$ for the polygonal path $\gamma$. Hence, the
center variational principle ensures that the phase is stationary
for the classical trajectories centered on $x$, then yielding 
\begin{equation}
U^{t}(x)_{SC}=\sum_{\gamma}\frac{2^{L}\exp\left\{ i\hbar^{-1}S_{\gamma}^{t}(x)+i\frac{\pi}{2}\alpha_{\gamma}^{t}\right\} }{|\det(\mathcal{M}^{t}+1)|{}^{\frac{1}{2}}}.\label{eq:USC-1}
\end{equation}
 The summation is over all the classical orbits $\gamma$ whose center
lies on the point $x$ after having evolved a time $t$ \cite{ozrep}.
Then $S_{\gamma}^{t}(x)$ is the classical center generating function
of the orbit, while $\mathcal{M}^{t}=\frac{\partial^{2}S_{\gamma}^{t}(x)}{\partial x^{2}}$
stand for the monodromy matrix and $\alpha_{\gamma}^{t}$ its \textit{Morse
index}.

The metaplectic operators form a \textquotedbl{}double covering\textquotedbl{}
of the symplectic matrices, since this property gives contributions
to the Morse index \cite{mehlig}. If we follow the evolution of the
symplectic matrix as the trajectory evolves, each time $\mathcal{M}^{t}$
crosses a manifold where $\det(\mathcal{M}^{t}+1)=0$ (caustic) the
path contribution undergoes a divergence changing the sign from $-\infty$
to $\infty$. This change of the sign lets the quantum phase proceed
by $\frac{\pi}{2}$. The Morse index $\alpha_{\gamma}^{t}$ therefore
changes by $\pm1$ when crossing caustics \cite{berry,gutz2}.

For sufficiently short times such that the variational problem has
an unique solution there will have a single chord. Although for longer
times, there will be bifurcations producing more chords. In the case
of a single orbit, the corresponding \textit{Morse index} $\alpha_{\gamma}^{t}=0$. 

As was shown in \cite{ozrep}, in order to obtain the coordinates
representation of the semiclassical propagator from the semiclassical
propagator in the Weyl representation (\ref{eq:USC-1}), a Fourier
transformation must be performed leading to the Van Vleck propagator.
However, the Weyl propagator of equation (\ref{eq:USC-1}) suffers
for caustic singularities whenever $|\det(\mathcal{M}^{t}+1)|=0$,
while the Van Vleck propagator has caustic singularities for other
points in phase space, (namely when $\frac{\partial^{2}S_{\gamma}^{t}(q,q')}{\partial q\partial q'}=0$,
with $S_{\gamma}^{t}(q,q')$ the action of the orbit $\gamma$ that
is a type I generating function for the associated canonical transformation. 

In order to obtain a coherent state path integral, expression (\ref{eq:UxPath})
is inserted in the coherent state propagator (\ref{eq:XUgaussX})
so that 
\begin{eqnarray}
 & \langle X_{1}|\widehat{U}^{t}|X_{2}\rangle\!=\!\lim_{k\rightarrow\infty}\!\int\!\frac{dx_{i}^{k}}{(\pi\hbar)^{kl}}e^{\frac{i}{\hbar}\left\{ \frac{X_{1}\wedge X_{2}}{2}-\frac{t}{k}\sum_{n=1}^{k}H(x_{n})\right\} }\nonumber \\
 & \!\!\!\times\!\int\frac{dx}{(\hbar\pi)^{l}}e{}^{\left\{ \frac{i}{\hbar}\left[x\wedge\xi_{0}+\Delta_{k+1}(x,x_{1},\!\cdots\!,x_{k}\!)\right]-\frac{\left(X-x\right)^{2}}{\hbar}\right\} }.\label{eq:Uxxxx}
\end{eqnarray}
 Recalling the linear relation of $\Delta_{k+1}$ with $x$, 
\begin{equation}
\!\Delta_{k+1}\!(\! x,x_{1},\!\cdots\!,x_{k}\!)=\! C+\!\xi\wedge x,\;\textrm{with}\;\:\xi\!=\!\sum_{j=1}^{k}(-1)^{j}x_{j},\label{eq:DELSIG}
\end{equation}
 the $x$ integral in (\ref{eq:Uxxxx}) is quadratic and can be solved
exactly, yielding 
\begin{eqnarray}
 & \langle X_{1}|\widehat{U}^{t}|X_{2}\rangle=e^{-\frac{iX_{1}\wedge X_{2}}{2\hbar}}\lim_{k\rightarrow\infty}\int\frac{dx_{i}^{k}}{(\pi\hbar)^{kl}}e^{-\frac{1}{4\hbar}\left(\bar{\xi}-\xi_{0}\right)^{2}}\nonumber \\
 & \times e^{\frac{i}{\hbar}\left\{ \Delta_{k+1}(X,x_{1},\cdots x_{k})-\frac{t}{k}\sum_{n=1}^{k}H(x_{n})\right\} }.\label{eq:XUXpath2}
\end{eqnarray}
 Here $\bar{\xi}$ is the chord passing through the mid point $X$.
Note that, according to (\ref{eq:DELSIG}) the chord $\bar{\xi}$
only depends on the other centers $x_{1},\cdots,x_{k}$. The usual
semiclassical limit of the CS propagator is obtained, in the limit
where $\hbar\rightarrow0$, with SPA in (\ref{eq:XUXpath2}). The
phase in (\ref{eq:XUXpath2}) is the center generating function (\ref{eq:CGF}).
According to the center variational principle the path integral is
stationary for classical orbits. Hence, analogously to (\ref{eq:USC-1}),
a SPA in (\ref{eq:XUXpath2}) yields 
\begin{equation}
\langle X_{1}\!|\widehat{U}^{t}\!|X_{2}\rangle_{SC1}\!=\!\sum_{\gamma}\!\frac{2^{l}e^{\left\{ \frac{i}{\hbar}\left[\widetilde{S}_{\gamma}^{t}(X)-\frac{1}{2}X_{1}\wedge X_{2}\right]-\frac{1}{4\hbar}\left(\xi_{\gamma}^{t}-\xi_{0}\right)^{2}\right\} }}{|\det({\cal M}_{\gamma}+1)|{}^{\frac{1}{2}}}\!.\!\label{eq:XUXSC1}
\end{equation}
 The sum runs now over all the classical orbits $\gamma$ whose center
lies on the point $X$, while $\xi_{\gamma}^{t}$ is the chord joining
the initial and final points of the orbit $\gamma$. For sufficiently
short times, there will have a single chord, while for longer times,
bifurcations will produce more chords. However, as a consequence of
a Gaussian cutoff on the length of the chords, the amplitude will
be severely damped if the classical chord $\xi_{\gamma}^{t}$ is long.
For the center generating function, $S_{\gamma}^{t}(x)$, we have
defined $\widetilde{S}_{\gamma}^{t}(x)=S_{\gamma}^{t}(x)+\hbar\frac{\pi}{2}\alpha_{\gamma}^{t}$
in order to include the Morse index in the action.

By means of an analytic continuation, dos Santos and de Aguiar have
obtained in \cite{aguiar} an expression similar to (\ref{eq:XUXSC1})
but involving complex trajectories. Although, with real trajectories,
expression (\ref{eq:XUXSC1}) does not give accurate results even
for quadratic Hamiltonian systems, as is shown in Figure 2. The SPA
used to go from (\ref{eq:XUXpath2}) to (\ref{eq:XUXSC1}) neglects
the Gaussian factor $(\bar{\xi}-\xi_{0})^{2}$, leading to a real
stationary trajectory. If the factor is taken into account, the exponent
becomes complex and the approximation would involve complex trajectories.
This approximation is also exact for quadratic Hamiltonians, but has
the drawback of dealing with complex trajectories.

Alternatively, to obtain a semiclassical approximation for the CS
propagator we can take advantage of the semiclassical approximation
for the propagator in the Weyl representation (\ref{eq:USC-1}). After
what, the transformation to CS representation is performed through
(\ref{eq:XUgaussX}) so that,
\begin{eqnarray}
\langle X_{1}\!\! & \!| & \!\widehat{U}_{SC}^{t}|X_{2}\rangle\!=\!\left(\frac{2}{\pi\hbar}\right)^{l}e^{\frac{i}{2\hbar}X_{1}\wedge X_{2}}\nonumber \\
\!\!\! & \!\!\!\times\!\! & \!\!\sum_{\gamma}\!\int\!\frac{\exp-\frac{1}{\hbar}\left(X-x\right)^{2}}{\left|\det\left({\cal M}_{\gamma}+1\right)\right|^{\frac{1}{2}}}e^{\frac{i}{\hbar}\left[\widetilde{S}_{\gamma}^{t}(x)+\xi_{0}\wedge x\right]}dx.\label{eq:XUscX}
\end{eqnarray}
 This procedure is equivalent to perform in (\ref{eq:Uxxxx}) a SPA
for the $dx_{i}^{k}$ variables before integrating in $dx$. 

In order to perform the $x$ integration in (\ref{eq:XUscX}) by a
usual SPA, the stationary phase point, $x_{0}$, must satisfies
\[
\xi(x_{0})=-\left.{\cal J}\frac{\partial S_{\gamma}^{t}(x)}{\partial x}\right|_{x=x_{0}}=\xi_{0}.
\]
 That is, the chord $\xi(x_{0})$ through $x_{0}$ must be equal to
$\xi_{0}=X_{1}-X_{2}$, the chord that joins the points $X_{1}$ and
$X_{2}$. Hence, the semiclassical approximation for the propagator
in CS that is obtained by a SPA in (\ref{eq:XUscX}) is
\begin{equation}
\langle X_{1}|\widehat{U}^{t}|X_{2}\rangle_{SC2}=\sum_{\gamma_{0}}\frac{2^{l}e^{\left\{ \frac{i}{\hbar}\left[\widetilde{S}_{\gamma_{0}}^{t}(\xi_{0})+\frac{1}{2}X_{1}\wedge X_{2}\right]-\frac{1}{\hbar}\left(x_{0}-X\right)^{2}\right\} }}{|\det({\cal M}_{\gamma_{0}}-1)|{}^{\frac{1}{2}}}.\label{eq:XUXSC2}
\end{equation}
 The sum now runs over all the classical orbits $\gamma_{0}$ whose
chord is $\xi_{0}$. The chord action $\widetilde{S}_{\gamma_{0}}^{t}(\xi_{0})=\widetilde{S}_{\gamma_{0}}^{t}(x_{0})-\xi_{0}\wedge x_{0}$
is the Legendre transform of $\widetilde{S}_{\gamma_{0}}^{t}(x_{0})$
\cite{ozrep}. The expression (\ref{eq:XUXSC2}) is complementary
to equation (\ref{eq:XUXSC1}), while expressed in terms of chords
instead of centers. Although, as is the case for the chord action
\cite{ozrep}, the expression (\ref{eq:XUXSC2}) diverges for very
short times where the monodromy matrix becomes the identity. Also,
for obtaining (\ref{eq:XUXSC2}) it was assumed that the Gaussian
term in (\ref{eq:XUscX}) is smooth close to the stationary point.
However, this is not the case in the semicalssical limit, if $\hbar\rightarrow0$
then the width of the Gaussian tends to zero. For this reasons, it
is not surprising that expression (\ref{eq:XUXSC2}) fails, as is
shown in Figure 2.

\section{Semiclassical Coherent States propagator}

\label{sec3}

The phase space integral in (\ref{eq:XUscX}) must then be performed
avoiding the usual SPA. For this purpose, it must be noted that classical
orbits that starts near $X_{2}$ and ends up near $X_{1}$ will have
an important contribution in (\ref{eq:XUscX}). These orbits have
their center points close to $X$. Hence, let us expand the center
action up to quadratic terms near the mid point $X$, so that, 
\begin{equation}
S_{\gamma}^{t}(x)=S_{\gamma}^{t}(X)+\xi_{\gamma}^{t}\wedge x^{\prime}+x^{\prime}B_{t}x^{\prime}+O(x^{\prime3}),\label{action}
\end{equation}
 with $x^{\prime}=X-x$ . Where $S_{\gamma}^{t}(X)$ is the action
of the orbit through the point $X$ for which the chord $\xi_{\gamma}^{t}$
is

\[
\xi_{\gamma}^{t}=\left.-{\cal J}\frac{\partial S_{\gamma}^{t}(x)}{\partial x}\right|_{x=X},
\]
 while, the symmetric matrix $B_{t}$ is the Cayley representation
of the symplectic matrix $\mathcal{M}^{t}$ 
\begin{equation}
{\mathcal{J}}\mathit{B}_{t}=\frac{1-\mathcal{M}^{t}}{1+\mathcal{M}^{t}}=\frac{1}{2}\frac{\partial^{2}S_{\gamma}^{t}(x)}{\partial x^{2}}.\label{eq:JB}
\end{equation}

After the quadratic expansion of the action (\ref{action}) is inserted
in (\ref{eq:XUscX}) we get

\begin{eqnarray}
\langle X_{1}|\widehat{U}_{SC}^{t}|X_{2}\rangle & = & \left(\frac{2}{\pi\hbar}\right)^{L}\sum_{\gamma}\frac{I}{\left|\det\left(\mathcal{M}^{t}+1\right)\right|^{\frac{1}{2}}}\label{eq:UI}\\
 & \times & \exp\frac{i}{\hbar}\left[\widetilde{S}_{\gamma}^{t}(X)-\xi_{0}\wedge X+\frac{1}{2}X_{1}\wedge X_{2}\right],\nonumber 
\end{eqnarray}
 with 
\begin{equation}
I=\int e^{\frac{1}{\hbar}\left\{ -x^{\dagger}\mathrm{\mathbf{\mathit{C}}}x+i\left[x^{\dagger}\mathit{B}_{t}x+\left(\xi_{\gamma}^{t}-\xi_{0}\right)\wedge x\right]\right\} }dx
\end{equation}
 a quadratic integral. The matrix $\mathbf{\mathrm{\mathbf{\mathit{C}}}}$
is the quadratic form that denotes the scalar product, 
\[
x^{\prime2}=x^{\prime}.x^{\prime}=x^{\prime\dagger}\mathrm{\mathbf{\mathit{C}}}x^{\prime},
\]
 where $x^{\dagger}$ stands for the transposed vector. Note that
in an orthonormal basis the matrix $\mathbf{\mathrm{\mathbf{\mathit{C}}}}$is
the identity.

We now perform exactly the quadratic integral, using

\begin{eqnarray}
I & = & \int\exp\left\{ -\frac{1}{\hbar}x^{\prime\dagger}\mathcal{V}_{t}x^{\prime}+\frac{1}{\hbar}Y.x^{\prime}\right\} dx^{\prime}\nonumber \\
 & = & \frac{\left(\pi\hbar\right)^{L}}{\sqrt{\left(\det\mathcal{V}_{t}\right)}}\exp\left\{ \frac{1}{4\hbar}Y^{\dagger}\mathcal{V}_{t}^{-1}Y\right\} .\label{eq:IYY-1}
\end{eqnarray}
 From equation (\ref{eq:UI}) 
\begin{equation}
\mathcal{V}_{t}=\mathbf{\mathit{C}}-i\mathit{B}_{t}\label{eq: VCB}
\end{equation}
 and 
\begin{equation}
Y=i{\cal J}\left(\xi_{\gamma}^{t}-\xi_{0}\right)=2i{\cal J}\delta_{\gamma}^{t},\label{eq:Ydelta}
\end{equation}
 where $\xi_{\gamma}^{t}=x_{f}-x_{i}$ is the chord that joins $x_{f}$
and $x_{i}$, respectively the final and initial point of the orbit
$\gamma$ of center $X$. This last expression defines the point shift
$\delta_{\gamma}^{t}$, so that 
\begin{equation}
\delta_{\gamma}^{t}=\frac{1}{2}\left(\xi_{\gamma}^{t}-\xi_{0}\right)=x_{f}-X_{1}=X_{2}-x_{i}.\label{eq:deltadef}
\end{equation}
 Note that, the point shift $\delta_{\gamma}^{t}$ is zero if there
is a classical orbit starting in the point $X_{2}$ and ending in
$X_{1}$. 

Inserting (\ref{eq:IYY-1}) in (\ref{eq:UI}), we get for the propagator
in coherent states, 
\begin{eqnarray}
\langle X_{1}|\widehat{U}_{SC}^{t}|X_{2}\rangle & = & 2^{L}\sum_{\gamma}\frac{\exp\left[\frac{-1}{\hbar}\delta_{\gamma}^{t\dagger}\mathcal{\widetilde{V}}\delta_{\gamma}^{t}\right]}{\left[\det\mathcal{V}_{t}\left|\det\left(\mathcal{M}^{t}+1\right)\right|\right]^{\frac{1}{2}}}\times\nonumber \\
 &  & \exp\frac{i}{\hbar}\left[\widetilde{S}_{\gamma}^{t}(X)-\frac{1}{2}X_{1}\wedge X_{2}\right]\label{eq:UYY}
\end{eqnarray}
 with the complex matrix $\mathcal{V}_{t}$ and the point shift $\delta_{\gamma}^{t}$
defined respectively in (\ref{eq: VCB}) and (\ref{eq:deltadef})
while $\mathcal{\widetilde{V}}={\mathcal{J}^{\dagger}}\mathcal{V}_{t}^{-1}{\mathcal{J}}$.
In order to separate amplitude and phase terms in (\ref{eq:UYY}),
it is useful to write

\begin{equation}
\mathcal{\widetilde{V}}={\mathcal{J}^{\dagger}}\mathcal{V}_{t}^{-1}{\mathcal{J}}={\mathcal{J}^{\dagger}}\frac{1}{\mathit{C}-i\mathit{B}_{t}}{\mathcal{J}}=\mathcal{\overline{C}}_{t}-i\mathcal{\overline{B}}_{t},\label{eq:VCB1}
\end{equation}
 with the real matrices
\[
\mathcal{\overline{C}}_{t}=\Re(\mathcal{\widetilde{V}})\quad\textrm{and}\quad\mathcal{\overline{B}}_{t}=-\Im(\mathcal{\widetilde{V}}).
\]
 Also,
\begin{equation}
\det\mathcal{V}_{t}=\biggl|\det\mathcal{V}_{t}\biggr|e^{i\varepsilon},\label{eq:detV}
\end{equation}
 with $\biggl|\det\mathcal{V}_{t}\biggr|$ denoting the modulus and
$\varepsilon$ the argument.

Hence inserting (\ref{eq:VCB1}) and (\ref{eq:detV}) in the~matrix
elements of the coherent state propagator (\ref{eq:UYY}) we obtain
\begin{eqnarray}
 & \langle X_{1}|\widehat{U}^{t}|X_{2}\rangle_{SC3}=\sum_{\gamma}\frac{2^{l}}{\left|\det\left[\mathcal{V}_{t}\left(\mathcal{M}^{t}+1\right)\right]\right|^{\frac{1}{2}}}\exp\left[-\frac{1}{\hbar}\delta_{\gamma}^{t\dagger}\mathcal{\overline{C}}_{t}\delta_{\gamma}^{t}\right]\nonumber \\
 & \times\exp\frac{i}{\hbar}\left[-\frac{1}{2}X_{1}\wedge X_{2}+\widetilde{S}_{\gamma}^{t}(X)+\delta_{\gamma}^{t\dagger}\mathcal{\overline{B}}_{t}\delta_{\gamma}^{t}+\hbar\frac{\varepsilon}{2}\right].\label{eq: XUXSC3}
\end{eqnarray}
The sum in (\ref{eq: XUXSC3}) runs over all the classical orbits
$\gamma$ whose center lies on the point $X$. The point shift $\delta_{\gamma}^{t}=\nicefrac{1}{2}\mathrm{\left(\xi_{\gamma}^{t}-\xi_{0}\right)}$
is half the difference between the chord $\xi_{\gamma}^{t}$, that
joins the initial and final points of the orbit $\gamma$, and the
chord $\xi_{0}$ joining the points $X_{1}$ and $X_{2}$ (see Figure
1).

The expression (\ref{eq: XUXSC3}) of the semiclassical matrix elements
of the quantum propagator between two CS, is the main contribution
of this work. It is entirely expressed in terms of real classical
objects, namely the action $S_{\gamma}^{t}(X)$ of the classical real
orbit whose mid point is $X$, the point shift $\delta_{\gamma}^{t}$,
the monodromy matrix $\mathcal{M}^{t}$ and its Cayley representation
$B_{t}$.

Very importantly, let us see that,
\begin{eqnarray*}
\mathcal{V}_{t}\left(\mathcal{M}^{t}+1\right) & = & \left(\mathbf{\mathit{C}}-i\mathit{B}_{t}\right)\left(\mathcal{M}^{t}+1\right).\\
 & = & C\left(\mathcal{M}^{t}+1\right)+i{\mathcal{J}}\left(1-\mathcal{M}^{t}\right),
\end{eqnarray*}
 hence, the only way for the pre-exponential factor in expression
(\ref{eq: XUXSC3}) to vanish ( i.e, $\det\left[\mathcal{V}_{t}\left(\mathcal{M}^{t}+1\right)\right]=0$)
is that the symplectic matrix $\mathcal{M}^{t}$ has simultaneously
eigenvalues $+1$ and $-1$. This is never the case for a one degree
of freedom system, while for systems with two and more degrees of
freedom it is an accidental coincidence of crossing of two types of
caustics.

It is also important to remark that, the Gaussian term in (\ref{eq: XUXSC3})
dampens the amplitude for large values of the point shift $\delta_{\gamma}^{t}$,
that is for orbits that start far from the point $X_{2}$ (or end
far from $X_{1}$). So, the main contribution in the sum over classical
orbits in (\ref{eq: XUXSC3}) will come from the single orbit $\gamma$
whose initial point lies the closest to $X_{2}$. Then, only this
particular orbit will be taken into account.

It must be mentioned that the same expression (\ref{eq: XUXSC3})
could have been obtained by performing in (\ref{eq:XUXpath2}) the
complete quadratic integral instead of the SPA that led to (\ref{eq:XUXSC1}).

Also, note that, if $t=0$, the quantum propagator is just the identity
operator in Hilbert space, the classical symplectic matrix is the
identity, the center action is null, and so are the symmetric matrix
$\left(\mathit{B}_{t=0}=0\right)$ and the chord $\xi_{\gamma}^{t}=2\delta_{\gamma}^{t}-X_{2}+X_{1}=0$
. Hence we recover the result (\ref{cohcoh}) for the overlap of coherent
states.

In expression (\ref{eq: XUXSC3}) the orbit $\gamma$ whose center
is $X$ remains to be determined in order to obtain its action $\widetilde{S}_{\gamma}^{t}(X)$
and the point shift $\delta_{\gamma}^{t}$. As we only need the contribution
for the single orbit $\gamma$ whose initial point lies the closest
to $X_{2}$, we linearize the flux in the the neighborhood of the
orbit $\gamma_{2}$ passing through the point $X_{2}$ (see Figure
1). For that purpose, we use the center generating function, so that
\begin{equation}
S_{\gamma}^{t}(x)=S_{\gamma_{2}}^{t}+\xi_{\gamma_{2}}^{t}\wedge x_{2}+x_{2}^{\dagger}\mathit{B}_{t}x_{2}+O(x^{\prime3}),\label{action-1}
\end{equation}
 with $x_{2}=x-X_{\gamma_{2}}$, where $X_{\gamma_{2}}$ denotes the
mid point of this orbit $\gamma_{2}$ starting in $X_{2}$ and ending
up in $x_{2f}$. The chord $\xi_{\gamma_{2}}^{t}=-{\cal J}\frac{\partial S_{\gamma}^{t}(x)}{\partial x}\biggr|_{X_{2}}=x_{2f}-X_{2}$
joins its end points, while $S_{\gamma_{2}}^{t}$ stands for the action
of the orbit. Note that $X_{\gamma_{2}}=X_{2}+\nicefrac{1}{2}\xi_{\gamma_{2}}^{t}$.

\begin{figure}
\includegraphics[width=11cm]{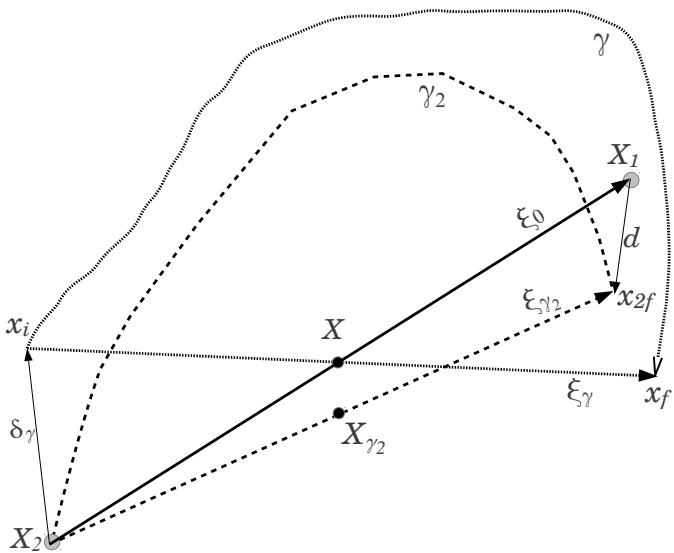}

\caption{The points $X_{2}$ and $X_{1}$ have their mid point in $X$ and
$\xi_{0}$ denotes the chord that joins them. The orbit $\gamma$,
has also its center point in $X$ while its chord $\xi_{\gamma}^{t}$
joins the points $x_{i}$ and $x_{f}$. The orbit $\gamma$, starts
in $x_{i}$ that is shifted from $X_{2}$ by the point shift $\delta_{\gamma}^{t}$.
Meanwhile, orbit $\gamma_{2}$ starts in $X_{2}$ and evolves after
a time $t$ up to $x_{2f}$, the drift from $X_{1}$ is denoted by
$d^{t}$. For this last orbit $\gamma_{2}$, the chord is denoted
by $\xi_{\gamma_{2}}^{t}$, while the mid point is $X_{\gamma_{2}}$.
For simplicity, in the figure we have dropped the time superscripts
$t$. }
\end{figure}

The chord $\xi_{\gamma}^{t}$ of the orbit $\gamma$ centered in $X$
is obtained by performing the derivative of the center generating
function (\ref{action-1}), 
\[
\xi_{\gamma}^{t}=-{\cal J}\frac{\partial S_{\gamma}^{t}(x)}{\partial x}\biggr|_{X}=\xi_{\gamma_{2}}^{t}-{\cal J}\mathit{B}_{t}\left(\xi_{0}-\xi_{\gamma_{2}}^{t}\right).
\]
 So that, recalling (\ref{eq:JB}) , we get for the point shift 
\begin{equation}
\delta_{\gamma}^{t}=\frac{1}{2}\left(\xi_{\gamma}^{t}-\xi_{0}\right)=\frac{1}{\mathcal{M}^{t}+1}d^{t},\label{eq:deltaXX}
\end{equation}
 where, 
\begin{equation}
d^{t}=\xi_{\gamma_{2}}^{t}-\xi_{0}=x_{2f}-X_{1}\label{eq:d2}
\end{equation}
 is the drift of the trajectory $\gamma_{2}$ that starting in $X_{2}$
ends up in $x_{2f}$ instead of $X_{1}$ (see Figure 1). In order
to include the Morse index~$\alpha_{\gamma_{2}}^{t}$ in the action
let us define the action $\widetilde{S}_{\gamma_{2}}^{t}=S_{\gamma_{2}}^{t}+\hbar\frac{\pi}{2}\alpha_{\gamma_{2}}^{t}$.
Hence, for the center action of the orbit $\gamma$ whose middle point
is $X$ is we get
\begin{eqnarray}
\widetilde{S}_{\gamma}^{t}(X) & = & \widetilde{S}_{\gamma_{2}}^{t}+\frac{1}{2}\xi_{\gamma_{2}}^{t}\wedge\xi_{0}+\frac{1}{4}d^{t\dagger}\mathit{B}_{t}d^{t}.\label{eq:SX}
\end{eqnarray}
With expressions (\ref{eq:SX}) and (\ref{eq:deltaXX}), respectively
for the point shifts $\delta_{\gamma}^{t}$ and the center action
$\widetilde{S}_{\gamma}^{t}(X)$, inserted in (\ref{eq: XUXSC3})
we obtain for the coherent states propagator,
\begin{align}
 & \langle X_{1}|\widehat{U}^{t}|X_{2}\rangle_{SC3}=\frac{2^{l}}{\left|\det\left[\mathcal{V}_{t}\left({\cal M}^{t}+1\right)\right]\right|^{\frac{1}{2}}}\exp\left[-\frac{d^{t\dagger}E_{t}d^{t}}{\hbar}\right]\nonumber \\
 & \times\exp\frac{i}{\hbar}\left[\mathcal{S}_{2}^{t}+\left(\xi_{\gamma_{2}}^{t}+X\right)\wedge\frac{\xi_{0}}{2}+d^{t\dagger}\left(\frac{\mathit{B}_{t}}{4}+D_{t}\right)d^{t}\right].\label{eq:XUXSCLIN}
\end{align}
Now $\varepsilon$ , the argument of $\det\left[\mathcal{V}_{t}\right]$
has been included in the action
\[
\mathcal{S}_{2}^{t}=\widetilde{S}_{\gamma_{2}}^{t}+\hbar\frac{\varepsilon}{2}=S_{\gamma_{2}}^{t}+\hbar\frac{\pi}{2}\alpha_{\gamma_{2}}^{t}+\hbar\frac{\varepsilon}{2}
\]
and the symmetric matrices $D_{t}$ and $E_{t}$ are defined as
\begin{eqnarray}
D_{t} & = & \left(\frac{1}{\mathcal{M}^{t}+1}\right)^{\dagger}\mathcal{\overline{B}}_{t}\left(\frac{1}{\mathcal{M}^{t}+1}\right),\nonumber \\
E_{t} & = & \left(\frac{1}{\mathcal{M}^{t}+1}\right)^{\dagger}\mathcal{\overline{C}}_{t}\left(\frac{1}{\mathcal{M}^{t}+1}\right)\textrm{ }.\label{eq:DyE}
\end{eqnarray}

Equation (\ref{eq:XUXSCLIN}) is a general expression only in term
of classical objects, its difference from (\ref{eq: XUXSC3}) is that
we have made use of the quadratic expansion of the action around the
orbit $\gamma_{2}$ in order to express both the point shift $\delta_{\gamma}^{t}$
and the center generating function $\widetilde{S}_{\gamma}^{t}(X)$
of the orbit $\gamma$ only in terms of magnitudes given by the orbit
$\gamma_{2}$ passing through $X_{2}$. This is a crucial advantage,
now the semiclassical approximation of the CS matrix elements involves
uniquely objects obtained directly by the classical evolution of the
flux through $X_{2}$, namely, the chord $\xi_{\gamma_{2}}^{t}$ (or
the drift $d^{t}$ defined in (\ref{eq:d2})), and the action $\mathcal{S}_{2}^{t}$
of the classical orbit $\gamma_{2}$ passing thorough $X_{2}$, and
the monodromy matrix $\mathcal{M}^{t}$. From this former, equation
(\ref{eq:JB}) gives its Cayley representation $\mathit{B}_{t}$,
after what, with (\ref{eq: VCB}), we get the complex matrix $\mathcal{V}_{t}$
while the real matrices $\mathcal{\overline{C}}_{t}$ and $\mathcal{\overline{B}}_{t}$
defined in (\ref{eq:VCB1}) allows to obtain $D_{t}$ and $E_{t}$
through (\ref{eq:DyE}). Hence, there is no need for any further root
search, neither integration over phase space conditions.

It is very useful to perform a comparison of the here derived Semiclassical
CS propagator with other kinds of propagators based on Gaussian wave
packets, namely the initial value representations (IVR) of the propagator.
These propagators are generally based on wave packets of the form
\[
\langle q|X^{\tau}\rangle=\left(\frac{\Re\tau}{\pi}\right)^{\frac{1}{4}}\exp\!\left[\!-\frac{\tau}{2}(q-Q)^{2}\!+\! i\frac{P}{\hbar}\left(q\!-Q\right)\!\right],
\]
which resembles (\ref{eq:qX}) if $\tau=\frac{m\omega}{\hbar}$. Although,
the different phase factor ensures the symplectic invariance of the
coherent states used in (\ref{eq:qX}). In order to maintain the symplectic
form we will then chose as in \cite{DE},
\begin{equation}
\langle q|X^{\tau}\!\rangle\!=\!\left(\frac{\Re\tau}{\pi}\right)^{\frac{1}{4}}\!\!\exp\!\left[\!-\frac{\tau}{2}(q-Q)^{2}\!+\! i\frac{P}{\hbar}\left(q\!-\frac{Q}{2}\right)\!\right]\!.\!\label{eq:qXg}
\end{equation}
 With this choice, the overlap between two Gaussian wave packets is:
\begin{equation}
\langle X_{1}^{\tau_{1}}|X_{2}^{\tau_{2}}\rangle\!=\!\exp\!\left[-\frac{\left(X_{1}-X_{2}\right)^{2}}{4\hbar}\!-\!\frac{i}{2\hbar}X_{1}\wedge X_{2}\!\right],\label{eq:XX}
\end{equation}
 the norm is defined as $X^{2}=X^{\dagger}\Lambda X$ where the symplectic
squeezing matrix is 
\[
\Lambda=\left(\begin{array}{cc}
\left(\hbar\sqrt{\tau_{1}\tau_{2}}\right)^{-1} & 0\\
0 & \hbar\sqrt{\tau_{1}\tau_{2}}
\end{array}\right).
\]

The IVR of the propagator, in coordinates representation takes the
form \cite{kay94}:
\begin{eqnarray}
\!\!\! K^{t}(q,q')\! & \!\!=\! & \langle q|\widehat{U}^{t}|q'\rangle_{\! I\! V\! R\!}\nonumber \\
\!\! & \!\!\!\!=\!\!\!\! & \!\!\!\int\!\!\!\frac{dX_{0}}{\left(2\pi\hbar\right)^{l}}\langle q|X_{t}^{\tau_{1}}\!\rangle R_{t}(\! X_{0}\!)e^{\frac{i}{\hbar}S^{t}\!(\! X_{0}\!)}\langle X_{0}^{\tau_{2}}\!|q'\rangle,\label{eq:KHK}
\end{eqnarray}
where $\langle X_{0}^{\tau_{2}}|q'\rangle$ is a Gaussian wave packet
whose center lies in the point $X_{0}$ (initial point) in phase space.
Meanwhile $\langle q|X_{t}^{\tau_{1}}\rangle$ is a Gaussian wave
packet centered in the point $X_{t}$, that is the classically evolved
initial point $X_{0}$ up to a time $t$. For the pre-exponential
factor,
\begin{equation}
\!\! R_{t}(\! X_{0}\!)\!=\!\!\left[\! A\!\left(\!\!\tau_{1}m_{11}\!+\!\tau_{2}m_{22}\!-\! i\hbar\tau_{1}\tau_{2}m_{21}\!-\!\frac{1}{i\hbar}m_{12}\!\!\right)\!\right]^{\frac{1}{2}}\!,\!\label{eq:RIVR}
\end{equation}
 where $A=\frac{1}{2\sqrt{\Re\tau_{1}\Re\tau_{2}}}$ and the monodromy
matrix elements
\[
\left(\begin{array}{c}
\delta Q_{t}\\
\delta P_{t}
\end{array}\right)=M\left(\begin{array}{c}
\delta Q_{0}\\
\delta P_{0}
\end{array}\right)=\left(\begin{array}{cc}
m_{11} & m_{12}\\
m_{21} & m_{22}
\end{array}\right)\left(\begin{array}{c}
\delta Q_{0}\\
\delta P_{0}
\end{array}\right)
\]
connects the initial and final deviations of the trajectories $X_{t}$.
While 
\[
S^{t}(X_{0})=\int_{0}^{t}\left[p{}_{t'}\dot{q}{}_{t'}-H\right]dt'
\]
 is the classical action of the orbit that starts in $X_{0}$. The
methods based on (\ref{eq:KHK}) are called initial value representation
(IVR) and have shown to be very useful for many physical systems.
They present the advantage over Van Vlek Gutwiller (VVG) \cite{VVleck,gutz2}
propagator that there is no need for any search of trajectories satisfying
special boundary conditions. 

Herman and Kluk (HK) formula is an IVR of the propagator that is also
known as Frozen Gaussian Approximation (FGA) since in that case, the
initial and final Gaussian have the same parameters $\tau_{1}=\tau_{2}=\tau$,
a real positive constant. With this choice for the Gaussian parameters,
the pre-factor (\ref{eq:RIVR}) for the HK propagator is 
\begin{equation}
\! R_{t}^{H\! K}\!(\! X_{0}\!)\!=\!\left|\!\frac{1}{2}\left(m_{11}\!+m_{22}-i\hbar\tau m_{21}-\frac{1}{i\hbar\tau}m_{12}\right)\!\right|\!^{\frac{1}{2}}\!,\label{eq:RHK}
\end{equation}
that never vanishes. Hence the HK propagator is free of caustics singularities.

An also well known case of IVR is Heller's Thawed Gaussian Approximation
(TGA), where now the initial and final Gaussian parameters differ.
While $\tau_{2}=\tau$ with again $\tau$ real positive,
\[
\tau_{1}=-\frac{i}{\hbar}\frac{m_{12}+i\hbar\tau m_{11}}{m_{22}+i\hbar\tau m_{21}}
\]
 is complex. The prefactor (\ref{eq:RIVR}) for this choice of parameters
has now the form
\begin{equation}
\! R_{t}^{T\! G\! A}\!(\! X_{0}\!)\!=\!\left(\frac{\tau}{\Re\tau_{1}}\right)^{\frac{1}{4}}\left(m_{22}+i\hbar\tau m_{21}\right)^{-\frac{1}{2}}.\label{eq:RTGA}
\end{equation}

The IVR propagator can also be expressed in a mixed representation,
\begin{eqnarray}
\!\!\!\! K^{t}(q,X_{2}) & \!\!\!= & \!\langle q|\widehat{U}^{t}|X_{2}\rangle_{I\! V\! R\!}=\int dq'K^{t}(q,q')\langle q'|X_{2}\rangle\!\nonumber \\
\!\!\!\! & \!\!\!\!\!\!\!\!=\!\!\!\! & \!\!\!\!\!\int\!\!\!\!\frac{dX_{0}}{\left(2\pi\hbar\right)^{l}}\!\langle q|\! X_{t}^{\tau_{1}}\rangle R_{t}\!(\! X_{0}\!)e^{\frac{i}{\hbar}S^{t}\!(\! X_{0}\!)}\langle X_{0}^{\tau_{2}}\!|X_{2}\rangle.\label{eq:HKmixto}
\end{eqnarray}
 Note that we omit the $\tau$ superscript for the special case where
$\tau=\frac{m\omega}{\hbar}$. The overlap between the wave functions
can be performed analytically using (\ref{eq:qXg}) and (\ref{eq:XX}),
whereas the integration over initial phase space is still left over
in the IVR propagator, but it is cut off by a bell shaped weight function
(overlap between two Gaussians). One can perform this integration
using Monte Carlo methods \cite{MC} leading to a powerful numerical
semiclassical procedure.

The HK propagator is a uniform semiclassical approximation of the
exact propagator as has been shown by Kay \cite{kay}. Indeed it is,
the lowest order term of an expansion of the propagator in $\hbar$.
Also, the HK propagator maintains unitarity for longer times than
other IVR such as Heller's TGA \cite{Harabati}.

Although, as was stated by Grossmann and Herman in \cite{GH}, a true
HK-like expression must consist of an integration over initial phase
space, which may not be treated in any additional approximation. However,
it has been shown by Grossmann in \cite{comment} that by a quadratic
expansion of the exponent around the phase space center of the initial
wavepacket, the TGA, originally derived by Heller in the mixed representation,
can be obtained from the HK propagator, equation (\ref{eq:HKmixto})
with (\ref{eq:RHK}) and $\tau_{1}=\tau_{2}=\tau$.

Also, Baranger et al \cite{Baranger} have shown that the TGA in the
mixed representation is equivalent to their mixed propagator obtained
with complex trajectories, except for two (related) differences. The
presence of an extra phase that is associated with the use of the
Gaussian averaged Hamiltonian $H$ in the computation of the action,
rather than the Weyl symbol $H(x)$ (essentially the classical Hamiltonian$H_{C}$).
In this respect, also Grossmann and Xavier \cite{GX} have derived
the HK propagator from the CS propagator proposed by Baranger el al.
restricting themselves to real variables.

For the IVR of the propagator in the CS representation,
\begin{eqnarray}
\!\!\!\! K^{t}(\! X_{1}\! & \!, & \! X_{2}\!)\!=\!\langle X_{1}|\widehat{U}^{t}|X_{2}\rangle_{I\! V\! R\!}\!=\!\int\!\! dq\langle X_{1}|q\rangle K^{t}(q,X_{2})\!\nonumber \\
\!\!\!\! & \!\!\!\!\!\!\!\!\!=\!\!\!\! & \!\!\!\!\!\int\!\!\!\!\frac{dX_{0}}{\left(2\pi\hbar\right)^{l}}\!\langle X_{1}\!|X_{t}^{\tau_{1}}\rangle R_{t}\!(\! X_{0}\!)e^{\frac{i}{\hbar}S^{t}\!(\! X_{0}\!)}\langle X_{0}^{\tau_{2}}\!|X_{2}\rangle\label{eq:KSC}
\end{eqnarray}
 the overlap between the Gaussians can be performed analytically,
whereas, the phase space integration must be done numerically without
any further approximation \cite{GH}.

In this CS representation, Deshpande and Ezra \cite{DE} found that
expanding the exponent of the integrand in (\ref{eq:KSC}) up to quadratic
terms and integrating , the linearized matrix element for the HK propagator
conditions, they obtained an expression that is identical with Littlejohn
form of the TGA matrix element \cite{little}. Also, this expression
resembles to the one obtained by Baranger et al \cite{Baranger} except
for the two differences previously described, that are related with
the use of the Gaussian averaged Hamiltonian $H$, rather than the
Weyl symbol $H(x)$.

A similar situation has been discussed by dos Santos and Aguiar \cite{aguiar},
in order to obtain a CS path integral in the Weyl representation,
who precisely argued the same difference with the representation used
by Barranger et al. in \cite{Baranger}. Indeed the expression obtained
from the linearized HK propagator by Deshpande and Ezra in \cite{DE}
coincides with the expression given by dos Santos and Aguiar in \cite{aguiar}.
However, Deshpande and Ezra \cite{DE} used real variables in their
derivation not complex ones. This means that the linearized version
of the HK propagator obtained in \cite{DE} gives the expression (\ref{eq:XUXSC1})
with real trajectories that is shown in section \ref{sec4} not to
give accurate results. Hence, as was stated by Grossmann and Herman
in \cite{GH}, a true HK-like expression must consist of an integration
over initial phase space, which may not be treated in any additional
approximation.

On the other side, the expression (\ref{eq: XUXSC3}) derived in this
work is a semiclassical approximation that has been obtained directly
from the semiclassical Weyl propagator. This last one, is the lowest
order term of an expansion in $\hbar$, obtained through SPA but of
the path integral expression of the propagator expressed in the Weyl
representation, that is symplectically invariant. Only afterwards,
the propagator is changed to the CS representation. For this last
procedure, we avoid SPA, instead, we performed a quadratic expansion
of the center action in the neighborhood of the relevant trajectories. 

Then, the obtained expression (\ref{eq: XUXSC3}) has the advantages
of being a symplectically invariant expression only dealing with real
trajectories. Also, differently from VVG propagators, it is free of
caustic singularities. Although, it does not need any phase space
integration, as is the case for IVR methods, the expression (\ref{eq: XUXSC3})
has the drawback for the need of searching trajectories whose center
lies in $X$. However, the quadratic expansion of the center action
and its use as a generating function allowed us to obtain expression
(\ref{eq:XUXSCLIN}) that only involves objects relative to the orbit
$\gamma_{2}$ passing though the initial point $X_{2}$.

\section{Matrix elements for maps and application to the cat map}

\label{sec4}

In what follows, we will obtain explicit expression for the classical
objects involved in (\ref{eq:XUXSCLIN}). In analogy with classical
Poincaré surfaces of section. We will first perform the study on a
surface of section that is transversal to the flux and passing through
$X$. The flux restricted to this section is now a map on the section,
for this map the time is discrete.

The study of autonomous fluxes through a map on surface of section
is a standard procedure, in the case of billiards this is done through
the well known Birkhoff coordinates. Also, quantum surface of section
methods are shown to be exact \cite{prosen} for general Hamiltonian
systems. From now on, the $2l$ dimensional autonomous flux is studied
through the $2l-2$ map on the mentioned surface of section.

As we have already mentioned, we need to evaluate the classical objects
involved in (\ref{eq:XUXSCLIN}). For that purpose, it will be convenient
to express them in the basis of eigenvectors of the symplectic matrix.
For the case of a map with one degree of freedom (corresponding to
a two degrees of freedom flux), this is the stable and unstable vector
basis $\left(\vec{\zeta_{u}},\vec{\zeta_{s}}\right)$ where the eigenvalues
of the symplectic matrix $\mathcal{M}^{t}$ are $\exp(-\lambda t)$
and $\exp(\lambda t)$, ($\lambda$ is the stability or Lyapunov exponent
of the orbit).

Let us then define $x_{s}$ and $x_{u}$ as canonical coordinates
along the stable and unstable directions respectively such that $x=(x_{u},x_{s})=x_{u}\vec{\zeta_{u}}+x_{s}\vec{\zeta_{s}}$
with $\vec{\zeta_{u}}\wedge\vec{\zeta_{s}}=1$. As the basis formed
by $\left(\vec{\zeta_{u}},\vec{\zeta_{s}}\right)$ is non orthonormal,
the scalar product of two vectors takes the form,
\begin{eqnarray*}
x.y & = & x^{\dagger}\mathrm{\mathit{C}}y\\
 & = & \left[\zeta_{u}^{2}x_{u}y_{u}+\zeta_{s}^{2}x_{s}y_{s}+\vec{\zeta_{u}}.\vec{\zeta_{s}}\left(x_{u}y_{s}+x_{s}y{}_{u}\right)\right].
\end{eqnarray*}
 That is, the scalar product matrix is,
\begin{equation}
\mathrm{\mathit{C}}=\left[\begin{array}{cc}
\zeta_{u}^{2} & \vec{\zeta_{u}}.\vec{\zeta_{s}}\\
\vec{\zeta_{u}}.\vec{\zeta_{s}} & \zeta_{s}^{2}
\end{array}\right]\label{X-xr}
\end{equation}
 with $\zeta_{u}^{2}=\vec{\zeta_{u}}.\vec{\zeta_{u}}$ and $\zeta_{s}^{2}=\vec{\zeta_{s}}.\vec{\zeta_{s}}$.
Since the transformation from the orthonormal basis $\left(\vec{i},\vec{j}\right)$
to the basis $\left(\vec{\zeta_{u}},\vec{\zeta_{s}}\right)$ is symplectic
\[
\det\mathit{C}=\zeta_{u}^{2}\zeta_{s}^{2}-\left(\vec{\zeta_{u}}.\vec{\zeta_{s}}\right)^{2}=1.
\]
 Also, in the $\left(\vec{\zeta_{u}},\vec{\zeta_{s}}\right)$ basis,
\begin{equation}
\mathcal{M}^{t}+1=2\cosh\left(\frac{\lambda t}{2}\right)\left[\begin{array}{cc}
e^{t\nicefrac{\lambda}{2}} & 0\\
0 & e^{-t\nicefrac{\lambda}{2}}
\end{array}\right],\label{eq:M+1}
\end{equation}
 hence
\begin{equation}
\left|\det\left(\mathcal{M}^{t}+1\right)\right|=4\cosh^{2}\left(\frac{\lambda t}{2}\right),\label{detM}
\end{equation}
 is easily obtained only in terms of $\lambda$ and $t$. Analogously,

\[
\mathcal{M}^{t}-1=2\sinh\left(\frac{\lambda t}{2}\right)\left[\begin{array}{cc}
e^{t\nicefrac{\lambda}{2}} & 0\\
0 & -e^{-t\nicefrac{\lambda}{2}}
\end{array}\right],
\]
 while, $\mathit{B}_{t}$, the Cayley parametrization of $\mathcal{M}^{t}$,
is in this basis 
\begin{equation}
\mathit{B}_{t}=\left[\begin{array}{cc}
0 & \tanh\left(t\lambda/2\right)\\
\tanh\left(t\lambda/2\right) & 0
\end{array}\right].\label{cayley}
\end{equation}
 Hence, using the expression of the symmetric matrix $\mathit{B}_{t}$
(\ref{cayley}) and the scalar product (\ref{X-xr}) we get the complex
matrix 
\begin{equation}
\!\mathcal{V}_{t}\!=\!\mathit{C}\!-\! i\mathit{B}_{t}\!=\!\!\left[\!\!\begin{array}{cc}
\zeta_{u}^{2} & \!\vec{\zeta_{u}}.\vec{\zeta_{s}}\!-\! i\tanh\left(\frac{t\lambda}{2}\right)\!\\
\!\vec{\zeta_{u}}.\vec{\zeta_{s}}\!-\! i\tanh\left(\frac{t\lambda}{2}\right)\! & \zeta_{s}^{2}
\end{array}\!\!\right]\!.\!\label{eq:Vlanda}
\end{equation}
 Also, the complex determinant 
\begin{equation}
\det\mathcal{V}_{t}=\left[1+\tanh^{2}\left(t\lambda/2\right)+2i\vec{\zeta_{u}}.\vec{\zeta_{s}}\tanh\left(t\lambda/2\right)\right],\label{eq:detv}
\end{equation}
 with modulus

\begin{equation}
\!\left|\det\mathcal{V}_{t}\right|\!=\!\sqrt{\!\left[\!1+\tanh^{2}\!\left(\!\frac{t\lambda}{2}\!\right)\!\right]^{2}\!+\!\left[\!2\vec{\zeta_{u}}.\vec{\zeta_{s}}\tanh\!\left(\!\frac{t\lambda}{2}\!\right)\!\right]^{2}}\!\label{eq:moddetv}
\end{equation}
 and argument 
\begin{equation}
\epsilon=\arctan\frac{2\vec{\zeta_{u}}.\vec{\zeta_{s}}\tanh\left(t\lambda/2\right)}{1+\tanh^{2}\left(t\lambda/2\right)},
\end{equation}
 can be explicitly written in terms of the time and the Lyapunov exponent.
Now, inverting the matrix $\mathcal{V}_{t}$ (\ref{eq:Vlanda}) we
get, 
\begin{eqnarray*}
\mathcal{V}_{t}^{-1} & = & \frac{-1}{\det\mathcal{V}_{t}}\left[\begin{array}{cc}
-\zeta_{s}^{2} & \vec{\zeta_{u}}.\vec{\zeta_{s}}\!-\! i\tanh\left(\frac{t\lambda}{2}\right)\\
\!\vec{\zeta_{u}}.\vec{\zeta_{s}}\!-\! i\tanh\left(\frac{t\lambda}{2}\right) & -\zeta_{u}^{2}
\end{array}\right]\\
 & = & \frac{1}{\det\mathcal{V}_{t}}\left(\mathit{C}^{-1}+i\mathit{B}_{t}\right).
\end{eqnarray*}
 We must note that since the matrix $\mathcal{V}_{t}$ is symmetric,
we get for the matrix $\mathcal{\widetilde{V}}$ defined in (\ref{eq:VCB1})
that,

\begin{eqnarray}
\mathcal{\widetilde{V}} & = & \frac{\mathcal{V}_{t}}{\det\mathcal{V}_{t}}=\frac{\left(\mathit{C}-i\mathit{B}_{t}\right)}{\left|\det\mathcal{V}_{t}\right|^{2}}\left(\Re(\mathbf{\det\mathcal{V}}_{t})-i\Im(\mathbf{\det\mathcal{V}}_{t})\right)\nonumber \\
 & = & \mathcal{\overline{C}}_{t}-i\mathcal{\overline{B}}_{t}.\label{eq:VCB}
\end{eqnarray}
 Hence, in the stable and unstable vector basis $\left(\vec{\zeta_{u}},\vec{\zeta_{s}}\right)$,
the real matrices $\mathcal{\overline{C}}_{t}$ and $\mathcal{\overline{B}}_{t}$
take the form

\begin{equation}
\mathcal{\overline{C}}_{t}=\frac{1}{\left|\det\mathcal{V}_{t}\right|^{2}}\left\{ \mathit{C}th_{2}-2\mathit{B}_{t}\vec{\zeta_{u}}.\vec{\zeta_{s}}\tanh\left(\frac{t\lambda}{2}\right)\right\} \label{eq:Cbar}
\end{equation}
 and

\begin{equation}
\mathcal{\overline{B}}_{t}=\frac{1}{\left|\det\mathcal{V}_{t}\right|^{2}}\left\{ \mathit{B}_{t}th_{2}+2\mathit{C}\vec{\zeta_{u}}.\vec{\zeta_{s}}\tanh\left(\frac{t\lambda}{2}\right)\right\} ,\label{eq:Bbar}
\end{equation}
 where $th_{2}=1+\tanh^{2}\left(\frac{t\lambda}{2}\right)$. The symmetric
matrix $\mathit{B}_{t}$, the scalar product matrix $\mathit{C}$
and the determinant $\det\mathcal{V}_{t}$ are respectively given
by the expressions (\ref{cayley}), (\ref{X-xr}) and (\ref{eq:moddetv}).
Inserting the expressions (\ref{eq:M+1}), (\ref{eq:Cbar}) and (\ref{eq:Bbar})
in the definition of the symmetric matrices $D_{t}$ and $E_{t}$
(\ref{eq:DyE}), we get 
\begin{equation}
D_{t}\!=\!\frac{-2\tanh\frac{t\lambda}{2}}{det_{1}}\!\left[\!\begin{array}{cc}
\!-\!\zeta_{s}^{2}\!\left(\!\vec{\zeta_{u}}.\vec{\zeta_{s}}\!\right)\! e^{-\! t\lambda} & th_{2}\!+\!2\!\left(\vec{\!\zeta_{u}}.\vec{\zeta_{s}}\!\right)\!^{2}\\
th_{2}\!+\!2\!\left(\vec{\!\zeta_{u}}.\vec{\zeta_{s}}\!\right)\!^{2} & -\zeta_{u}^{2}\!\left(\!\vec{\zeta_{u}}.\vec{\zeta_{s}}\!\right)\! e^{t\lambda}
\end{array}\!\right]\label{eq:Dlanda}
\end{equation}
 and

\begin{equation}
E_{t}=-\frac{th_{2}}{det_{1}}\left[\begin{array}{cc}
-\zeta_{s}^{2}e^{-t\nicefrac{\lambda}{2}} & \frac{\vec{\zeta_{u}}.\vec{\zeta_{s}}}{2\sinh^{2}\left(t\lambda/2\right)+1}\\
\frac{\vec{\zeta_{u}}.\vec{\zeta_{s}}}{2\sinh^{2}\left(t\lambda/2\right)+1} & -\zeta_{u}^{2}e^{t\nicefrac{\lambda}{2}}
\end{array}\right]\label{eq:Elanda}
\end{equation}
 where we have defined 
\[
det_{1}=4\cosh^{2}\left(\frac{t\lambda}{2}\right)\left|\det\mathcal{V}_{t}\right|^{2}.
\]
 It is important to note that, (\ref{eq:Dlanda}), (\ref{eq:Elanda}),
(\ref{cayley}), (\ref{eq:detv}) and (\ref{detM}) are respectively
explicit expression of the symmetric matrices $D_{t}$, $E_{t}$ and
$\mathit{B}_{t}$ and the determinants $\left|\det\mathcal{V}_{t}\right|$and
$\left|\det\left(\mathcal{M}^{t}+1\right)\right|$ for any value of
the time $t$. Inserting these expressions in (\ref{eq:XUXSCLIN}),
we obtain a semiclassical expression for the matrix elements of the
propagator in the CS basis entirely in terms of classical features
such as, the chord $\xi_{0}$ that joins the points $X_{2}$ and $X_{1}$,
the action of the orbit $\widetilde{S}_{X_{2}}$, the stable and unstable
vectors $\vec{\zeta_{u}},\vec{\zeta_{s}}$ and the Lyapunov exponent
$\lambda$.

Now the present theory is applied to the cat map i.e. the linear automorphism
on the $2$-torus generated by the $2\times2$ symplectic matrix $\mathcal{M}$,
that takes a point $x_{-}$ to a point $x_{+}$ : $x_{+}=\mathcal{M}x_{-}\quad\mbox{mod(1)}$.
In other words, there exists an integer $2$-dimensional vector $\mathbf{m}$
such that $x_{+}=\mathcal{M}x_{-}-\mathbf{m}$. Equivalently, the
map can also be studied in terms of the center generating function
\cite{mcat}. This is defined in terms of center points 
\begin{equation}
x\equiv\frac{x_{+}+x_{-}}{2}\label{xdef}
\end{equation}
 and chords 
\begin{equation}
\xi\equiv x_{+}-x_{-}=-\mathcal{J}\frac{\partial S(x,\mathbf{m})}{\partial x},\label{cordef}
\end{equation}
 where 
\begin{align}
S(x,\mathbf{m}) & =xBx+x\left(B-\mathcal{J}\right)\mathbf{m}+\frac{1}{4}\mathbf{m}(B+\widetilde{\mathcal{J}})\mathbf{m}\label{sx2}
\end{align}
 is the center generating function. Here $B$ is a symmetric matrix
(the Cayley parameterization of $\mathcal{M}$, as in (\ref{cayley})),
while 
\begin{equation}
\widetilde{\mathcal{J}}=\left[\begin{array}{c|c}
0 & 1\\
\hline 1 & 0
\end{array}\right].
\end{equation}
 We will study here the cat map with the symplectic matrix 
\begin{equation}
\mathcal{\! M}\!=\!\left[\!\begin{array}{cc}
2 & 3\\
1 & 2
\end{array}\!\right]\!\mbox{, and symmetric matrix }B\!=\!\left[\!\begin{array}{cc}
-\frac{1}{3} & 0\\
0 & 1
\end{array}\!\right]\!.\!\label{mhb}
\end{equation}
 This map is known to be chaotic, (ergodic and mixing) as all its
periodic orbits are hyperbolic. The map corresponds to viewing stroboscopically
the motion generated by a quadratic Hamiltonian \cite{key-7}. However,
the torus boundary conditions makes the dynamics as nonlinear as a
dynamics can get \cite{key-7}. The eigenvalues of $\mathcal{M}$
are $e^{-\lambda}$ and $e^{\lambda}$ with $\lambda=\ln(2+\sqrt{3})\approx1.317$.
This is then the stability exponent for the fixed points, whereas
the exponents must be doubled for orbits of period 2. All the eigenvectors
have directions $\vec{\zeta_{s}}=(-\frac{\sqrt{3}}{2},\frac{1}{2})$
and $\vec{\zeta_{u}}=(1,\frac{1}{\sqrt{3}})$ corresponding to the
stable and unstable directions respectively.

Quantum mechanics on the torus, implies a finite Hilbert space of
dimension $N=\frac{1}{2\pi\hbar}$, and that positions and momenta
are defined to have discrete values in a lattice of separation $\frac{1}{N}$
\cite{hanay,opetor}. The cat map was originally quantized by Hannay
and Berry \cite{hanay} in the coordinate representation the propagator
is: 
\begin{equation}
\langle\mathbf{q}_{k}|\hat{\mathbf{U}}_{\mathcal{M}}|\mathbf{q}_{j}\rangle=\left(\frac{i}{N}\right)^{\frac{1}{2}}{\exp}\left[\frac{i2\pi}{N}(k^{2}-jk+j^{2})\right],\label{uqq}
\end{equation}
 where the states $\langle q|\mathbf{q}_{j}\rangle$ are periodic
combs of Dirac delta distributions at positions $q=j/Nmod(1)$, with
$j$ integer in $[0,N-1]$. In the Weyl representation \cite{opetor},
the quantum map has been obtained in \cite{mcat} as 
\begin{align}
\mathbf{U}_{\mathcal{M}}(x) & =\frac{2}{\left|\det(\mathcal{M}+1)\right|^{\frac{1}{2}}}\sum_{\mathbf{m}}e^{i2\pi N\left[S(x,\mathbf{m})\right]}\label{ugxp}
\end{align}
 where the center points are represented by $x=(\frac{a}{N},\frac{b}{N})$
with $a$ and $b$ integer numbers in $[0,N-1]$ for odd values of
$N$ \cite{opetor}. There exists an alternative definition of the
torus Wigner function which also holds for even $N$.

The fact that the symplectic matrix $\mathcal{M}$ has equal diagonal
elements implies in the time reversal symmetry and then the symmetric
matrix $B$ has no off-diagonal elements. This property will be valid
for all the powers of the map and, using (\ref{ugxp}), we can see
that it implies in the quantum symmetry 
\begin{equation}
\mathbf{U}_{\mathcal{M}}^{l}(p,q)=\left(\mathbf{U}_{\mathcal{M}}^{l}(-p,q)\right)^{*}=\left(\mathbf{U}_{\mathcal{M}}^{l}(p,-q)\right)^{*}.\label{qsym}
\end{equation}
 for any integer value of $l$.

It has been shown \cite{hanay} that the unitary propagator is periodic
(nilpotent) in the sense that, for any value of $N$ there is an integer
$k(N)$ such that 
\[
\hat{\mathbf{U}}_{\mathcal{M}}^{k(N)}=e^{i\phi}.
\]
 Hence the eigenvalues of the map lie on the $k(N)$ possible sites
\begin{equation}
\left\{ \exp\left[\frac{i(2m\pi+\phi)}{k(N)}\right]\right\} ,\quad1\le m\le k(N).
\end{equation}
 For the cases where $k(N)\langle N$ there are degeneracies and the
spectrum does not behave as expected for chaotic quantum systems.
In spite of the peculiarities in this map, a very weak nonlinear perturbations
of cat maps restores the universal behavior of non degenerate chaotic
quantum systems spectra \cite{matos}. Eckhardt \cite{Eckhardt} has
argued that typically the eigenfunctions of cat maps are random.

The coherent states propagator on the torus depends on the definition
of the periodic coherent state \cite{nonen}, with $\langle p\rangle=P$
and $\langle q\rangle=Q$. In accordance to (\ref{eq:qX}) 
\begin{equation}
\langle{\bf X}|\mathbf{q}_{k}\rangle=\sum_{j=-\infty}^{\infty}e^{-\frac{1}{\hbar}\left[iP(j+\frac{Q}{2}-\nicefrac{k}{N})+\frac{1}{2}(j+Q-\nicefrac{k}{N})^{2}\right]}.\label{35}
\end{equation}

In order to construct operators or functions on the torus we have
to periodize the construction. This is done merely using the recipe
\cite{opetor} that for any operator its Weyl representation on the
torus ${\bf A}(x)$ is obtained from is analogue in the plane $A(x)$
by 
\begin{align*}
{\bf A}(x)=\sum_{j=-\infty}^{\infty}\sum_{k=-\infty}^{\infty}(-1)^{2ja+2kb+jkN}A(x+\frac{(k,j)}{2}).
\end{align*}
 Indeed the construction on the torus from the plane is obtain in
terms of averages over equivalent points, that are obtained by translation
with integer chords: $\widehat{T}_{\overrightarrow{k}}$ where $\overrightarrow{k}=\left(k_{p},k_{q}\right)$
is a two dimensional vector with integer components $k_{p}$ and $k_{q}$.
Hence, the unit operator in the Hilbert space of the torus is \cite{opetor}

\[
\boldsymbol{\widehat{1}_{N}}=\sum_{k=0}^{N-1}|\mathbf{q}_{k}\rangle\langle\mathbf{q}_{k}|=\biggl\langle\widehat{T}_{\overrightarrow{k}}e^{i2\pi(\chi\wedge\overrightarrow{k}+\frac{N}{4}\overrightarrow{k}{\mathcal{\widetilde{J}}}\overrightarrow{k})}\biggr\rangle
\]
 so that 
\begin{eqnarray*}
|{\bf X}\rangle & = & \boldsymbol{\widehat{1}_{N}}|X\rangle=\biggl\langle e^{i2\pi(\frac{N}{4}\overrightarrow{k}{\mathcal{\widetilde{J}}}\overrightarrow{k})}\widehat{T}_{\overrightarrow{k}}|X\rangle\biggr\rangle\\
 & = & \biggl\langle e^{i\pi\frac{N}{2}\overrightarrow{k}{\mathcal{\widetilde{J}}}\overrightarrow{k}}e^{-\frac{i}{2\hbar}X\wedge\overrightarrow{k}}|X+\overrightarrow{k}\rangle\biggr\rangle.
\end{eqnarray*}
 In this way the coherent states matrix elements for any operator
on the torus are obtained through

\begin{equation}
\langle\mathbf{X}_{1}|\mathbf{\widehat{A}}|\mathbf{X}_{2}\rangle\!=\!\biggl\langle\! e^{i\pi\frac{N}{2}\overrightarrow{k}{\mathcal{\widetilde{J}}}\overrightarrow{k}}e^{-\frac{i}{2\hbar}X\wedge\overrightarrow{k}}\langle X_{1}|\widehat{A}|X_{2}\!+\!\overrightarrow{k}\rangle\!\biggr\rangle.\!\label{eq:Atoro}
\end{equation}

Figure 2 shows the relative error $E$ on the amplitude of the semiclassical
approximations $\langle X_{1}|\widehat{U}^{t}|X_{2}\rangle_{SC1}$
and $\langle X_{1}|\widehat{U}^{t}|X_{2}\rangle_{SC2}$ obtained respectively
in (\ref{eq:XUXSC1}) and (\ref{eq:XUXSC2}) (after taking in both
cases the torus periodization (\ref{eq:Atoro})) with respect to the
exact expression obtained with the quantum propagator (\ref{uqq})
on the CS (\ref{35}). As can be seen, neither (\ref{eq:XUXSC1})
nor (\ref{eq:XUXSC2}) are good approximations of the exact CS matrix
elements, giving errors in the amplitude of more than 10\% or that
highly grow with $N$ respectively. Meanwhile, we have verified that
$\langle X_{1}|\widehat{U}^{t}|X_{2}\rangle_{SC3}$, obtained with
(\ref{eq:XUXSCLIN}), is exact in this case, for both the amplitude
and the phase, as expected in a linear system. 
\begin{figure}[H]
\includegraphics[clip,width=8.5cm]{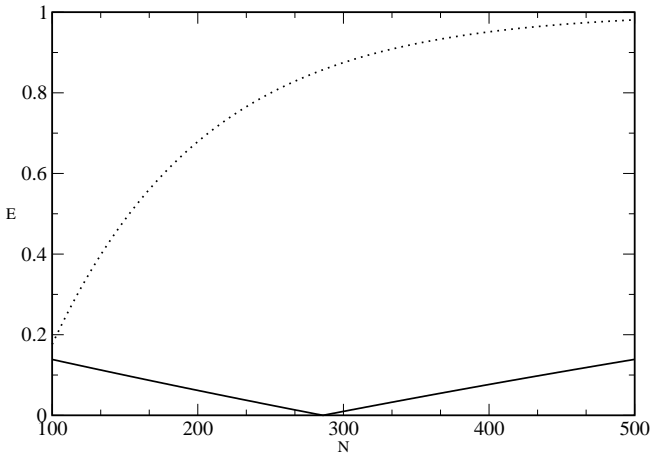} \caption{Relative error of the amplitude of the Semiclassical expressions $\langle X_{1}|\widehat{U}^{t}|X_{2}\rangle_{SC1}$
and $\langle X_{1}|\widehat{U}^{t}|X_{2}\rangle_{SC2}$ as a function
of $N$. In full line the error of $\langle X_{1}|\widehat{U}^{t}|X_{2}\rangle_{SC1}$
and in dotted line the error of $\langle X_{1}|\widehat{U}^{t}|X_{2}\rangle_{SC2}$.}
\end{figure}


\section{Conclusions}

\label{sec5}

To conclude, the expression (\ref{eq: XUXSC3}) obtained in this work
is an accurate semiclassical expression for the CS propagator that
avoids complex trajectories, it only involves real ones. For its obtainment
we have used the symplectically invariant Weyl representation. While
the, semiclassical Weyl propagator was derived by performing a SPA
for the path integral in the Weyl representation, for the transformation
to CS representation SPA was avoided.

Also, the quadratic expansion of the center generating function has
allowed to obtain a semiclassical expression of the CS propagator
(\ref{eq:XUXSCLIN}) involving only objects relative to the orbit
$\gamma_{2}$ passing though the initial point $X_{2}$, without the
need of any further search of trajectories that are typical procedures
of Gutzwiller Van Vleck based propagators \cite{gutz2,VVleck}, nor
phase space integration typical from IVR methods.

For the case of chaotic maps, the explicit time dependence of the
CS propagator has been derived only in terms of the action of the
orbit through the initial point $X_{2}$, the Lyapunov exponent and
the stable and unstable vector basis directions.

The comparison with a system whose semiclassical limit is exact has
allowed to correctly check the exactness of expression (\ref{eq:XUXSCLIN})
up to quadratic Hamiltonian systems.

It is important to mention that the present theory has already been
successfully applied for the semiclassical matrix elements for chaotic
propagators in the scar function basis \cite{scarelem}. This is a
crucial element in the semiclassical theory of short periodic orbits
for the evaluation of the energy spectrum of classically chaotic Hamiltonian
systems \cite{6ver}-\cite{13ver}.

Of course, the here derived expression can be applied to a vast variety
of systems, in particular for continuous Hamiltonian systems as was
done with complex trajectories in \cite{ribeiro}, indeed, the fact
that only real trajectories are involved guaranties a simpler procedure.

I thanks G. Carlo, E. Vergini and M. Saraceno for stimulating discussions
and the CONICET for financial support.


\appendix

\section*{Appendix: Reflection Operators in Phase Space}

Among the several representations of quantum mechanics, the Weyl-Wigner
representation is the one that performs a decomposition of the operators
that acts on the Hilbert space, on the basis formed by the set of
unitary reflection operators. In this appendix we review the definition
and some properties of this reflection operators.

First of all we construct the family of unitary operators 
\begin{equation}
\hat{T}_{q}=\exp(-i\hbar^{-1}q.\hat{p}),\qquad\hat{T}_{p}=\exp(i\hbar^{-1}p.\hat{q}),
\end{equation}
 and following \cite{ozrep}, we define the operator corresponding
to a general translation in phase space by $\xi=(p,q)$ as 
\begin{eqnarray}
\hat{T}_{\xi} & \equiv & \exp\left(\frac{i}{\hbar}\xi\wedge\hat{x}\right)\equiv\exp\left[\frac{i}{\hbar}(p.\hat{q}-q.\hat{p})\right]\\
 & = & \hat{T}_{p}\hat{T}_{q}\exp\left[-\frac{i}{2\hbar}\ p.q\right]\!=\!\hat{T}_{q}\hat{T}_{p}\exp\left[\frac{i}{2\hbar}\ p.q\right],\label{eq:tcor}
\end{eqnarray}
 where naturally $\hat{x}=(\hat{p},\hat{q})$. In other words, the
order of $\hat{T}_{p}$ and $\hat{T}_{q}$ affects only the overall
phase of the product, allowing us to define the translation as above.
$\hat{T}_{\xi}$ is also known as a \textit{Heisenberg operator}.
Acting on the Hilbert space we have: 
\begin{equation}
\widehat{T}_{\xi}|q_{a}\rangle=e^{\frac{i}{\hbar}p(q_{a}+\frac{q}{2})}|q_{a}+q\rangle\label{eq:tq}
\end{equation}
 and 
\begin{equation}
\widehat{T}_{\xi}|p_{a}\rangle=e^{-\frac{i}{\hbar}q(p_{a}+\frac{p}{2})}|p_{a}+p\rangle.
\end{equation}
 We, hence, verify their interpretation as translation operators in
phase space. The group property is maintained within a phase factor:
\begin{equation}
\hat{T}_{\xi_{2}}\hat{T}_{\xi_{1}}=\hat{T}_{\xi_{1}+\xi_{2}}e^{[\frac{-i}{2\hbar}\xi_{1}\wedge\xi_{2}]}=\hat{T}_{\xi_{1}+\xi_{2}}e^{[\frac{-i}{\hbar}D_{3}(\xi_{1},\xi_{2})]},\label{eq:tt}
\end{equation}
 where $D_{3}$ is the symplectic area of the triangle determined
by two of its sides. Evidently, the inverse of the unitary operator
$\hat{T}_{\xi}^{-1}=\hat{T}_{\xi}^{\dag}=\hat{T}_{-\xi}$ .

The set of operators corresponding to phase space reflections $\hat{R}_{x}$
about points $x=(p,q)$ in phase space, is formally defined in \cite{ozrep}
as the Fourier transform of the translation (or Heisenberg) operators
\begin{equation}
\widehat{R}_{x}\equiv(4\pi\hbar)^{-L}\int d\xi\quad e^{\frac{i}{\hbar}x\wedge\xi}\widehat{T}_{\xi}.\label{eq:rint}
\end{equation}
 Their action on the coordinate and momentum bases are 
\begin{eqnarray}
\hat{R}_{x}\left|q_{a}\right\rangle  & = & e^{2i(q-q_{a})p/\hbar}\;\left|2q-q_{a}\right\rangle \label{rqpl}\\
\hat{R}_{x}\left|p_{a}\right\rangle  & = & e^{2i(p-p_{a})q/\hbar}\;\left|2p-p_{a}\right\rangle ,
\end{eqnarray}
 displaying the interpretation of these operators as reflections in
phase space. Also, Using the coordinate representation of the coherent
state (\ref{eq:qX}) and the action of reflection on the coordinate
basis (\ref{rqpl}), we can see that the action of the reflection
operator $\widehat{R}_{x}$ on a coherent state $\left|X\right\rangle $
is the $x$ reflected coherent state 
\begin{equation}
\widehat{R}_{x}\left|X\right\rangle =\exp\left(\frac{i}{\hbar}X\wedge x\right)\left|2x-X\right\rangle .
\end{equation}

This family of operators have the property that they are a decomposition
of the unity (completeness relation) 
\begin{equation}
\hat{1}=\frac{1}{2\pi\hbar}\int dx\ \hat{R}_{x},\label{1R}
\end{equation}
 and also they are orthogonal in the sense that 
\begin{equation}
Tr\left[\hat{R}_{x_{1}}\ \hat{R}_{x_{2}}\right]=2\pi\hbar\;\delta(x_{2}-x_{1}).\label{trR}
\end{equation}
 Hence, an operator $\hat{A}$ can be decomposed in terms of reflection
operators as follows 
\begin{equation}
\hat{A}=\frac{1}{2\pi\hbar}\int dx\ A_{W}(x)\ \hat{R}_{x}.\label{rep}
\end{equation}
 With this decomposition, the operator $\hat{A}$ is mapped on a function
$A_{W}(x)$ living in phase space, the so called Weyl-Wigner symbol
of the operator. Using (\ref{trR}) it is easy to show that $A_{W}(x)$
can be obtained by performing the following trace operation 
\[
A_{W}(x)=Tr\left[\hat{R}_{x}\ \hat{A}\right].
\]
 Of course, as it is shown in \cite{ozrep}, the Weyl symbol also
takes the usual expression in terms of matrix elements of $\hat{A}$
in coordinate representation 
\[
A_{W}(x)=\int\left\langle q-\frac{Q}{2}\right|\hat{A}\left|q+\frac{Q}{2}\right\rangle \exp\left(-\frac{i}{\hbar}pQ\right)dQ.
\]

It was also shown in \cite{ozrep} that reflection and translation
operators have the following composition properties 
\begin{equation}
\widehat{R}_{x}\widehat{T}_{\xi}=\widehat{R}_{x-\xi/2}e^{-\frac{i}{\hbar}x\wedge\xi}\ ,\label{eq:rt}
\end{equation}
\begin{equation}
\widehat{T}_{\xi}\widehat{R}_{x}=\widehat{R}_{x+\xi/2}e^{-\frac{i}{\hbar}x\wedge\xi}\ ,\label{eq:tr}
\end{equation}
\begin{equation}
\widehat{R}_{x_{1}}\widehat{R}_{x_{2}}=\widehat{T}_{2(x_{2}-x_{1})}e^{\frac{i}{\hbar}2x_{1}\wedge x_{2}}\label{eq:rr}
\end{equation}
 so that
\begin{equation}
\widehat{R}_{x}\widehat{R}_{x}=\widehat{1}\ .
\end{equation}
 Now using (\ref{eq:rr}) and (\ref{eq:tr}) we can compose three
reflections so that 
\begin{equation}
\widehat{R}_{x_{2}}\widehat{R}_{x}\widehat{R}_{x_{1}}=e^{\frac{i}{\hbar}\Delta_{3}(x_{2},x_{1},x)}\widehat{R}_{x_{2}-x+x_{1}}
\end{equation}
 where $\Delta_{3}(x_{2},x_{1},x)=2(x_{2}-x)\wedge(x_{1}-x)$ is the
area of the oriented triangle whose sides are centered on the points
$x_{2},x_{1}$ and $x$ respectively.



\begin{thebibliography}{10}
\bibitem{path} E. Schrodinger, Naturwiss. \textbf{14}, 664 (1926).

\bibitem{coherent}J.R. Klauder and B.S. Skagerstam, \textsl{in Coherent
}states\textsl{: Applications in Physics and Mathematical Physics},
(Singapore: Word Scientific, 1985).

\bibitem{klauderarx}J.R. Klauder, \textit{``The Feynman Path Integral:
An Historic Slice''} in a \textit{``Garden of Quanta''}, Eds. J.
Arafune et al (Word Scientific, Singapore, 2003) , pp 55-77.

\bibitem{herman} M.F. Herman and E. Kluk Chem. Phys. \textbf{91},
27 (1984).

\bibitem{miller} W.H. Miller, Mol. Phys.\textbf{100}, 397 (2002).

\bibitem{pollak} J. Tatchen and E. Pollak, J. Chem. Phys.\textbf{
130}, 041103 (2009).

\bibitem{kay} K.G. Kay, Chem. Phys.\textbf{ 322}, 3 (2006).

\bibitem{kay2}K.G. Kay, J. Chem. Phys.\textbf{ 132}, 244110 (2010).

\bibitem{Baranger}M. Baranger , M. A. M. de Aguiar , F. Keck, H.J.
Korsch and B. Schellhaaß, J. Phys. A: Math. Gen.\textbf{ 34}, 7227
(2001). See also \textit{ibid} \textbf{35}, 9493 (2002) and references
there in.

\bibitem{aguiar} L.C. dos Santos and M.A.M. de Aguiar, J. Phys. A:
Math. Gen.{\tiny {} }\textbf{39}, 13465 (2006).

\bibitem{pathPRL}J. H. Wilson and V. Galitski, Phys. Rev. Lett.\textbf{
106}, 110401 (2011).

\bibitem{gutz2} M.C. Gutzwiller in \textquotedbl{}\textit{Chaos and
Quantum Physics}\textquotedbl{}, Les Houches Session LII, pg.205-248
Ed: M.-J. Giannonni, A. Voros and J. Zinn-Justin (1989).

\bibitem{VVleck}J. H. V. Vleck, Proc. Natl. Acad. Sci. U.S.A. \textbf{14},
178 (1928).

\bibitem{ozrep} A.M. Ozorio de Almeida, Physics Report \textbf{295},
266 (1998).

\bibitem{mehlig} B. Mehlig and M. Wilkinson, Ann. Phys. (Leipzig)
\textbf{10}, 541-559 (2001).

\bibitem{berry} M.V. Berry, Proc. R. Soc. A \textbf{423}, 219 (1989).

\bibitem{kay94}K.G. Kay, J. Chem. Phys. \textbf{100,} 4377 (1994).

\bibitem{Harabati}C. Harabati, J.M. Rost and F. Grossmann, J. Chem.
Phys. \textbf{120}, 26 (2004).

\bibitem{GH} F. Grossmann and M.F. Herman, J. Phys. A: Math. Gen.
\textbf{35}, 9489 (2002).

\bibitem{comment} F. Grossmann, Comments At. Mol. Phys., \textbf{34},
141 (1999).

\bibitem{MC} E. Kluk, M.F. Herman and H.L. Davis, J. Chem. Phys.
\textbf{84,} 326 (1986).

\bibitem{GX}F. Grossmann and A.L. Xavier Jr., Phys. Lett. A \textbf{243},
243 (1998).

\bibitem{DE} S.A. Deshpande and G.S. Ezra, J. Phys. A: Math. Gen.
\textbf{39}, 5067 (2006).

\bibitem{little}R.G. Littlejohn, Phys. Rep. \textbf{138}, 193 (1986).

\bibitem{prosen} T. Prosen, J. Phys. A: Math. Gen. \textbf{28}, 4133
(1995).

\bibitem{mcat} A.M.F. Rivas, M. Saraceno and A.M. Ozorio de Almeida,
Nonlinearity \textbf{13}, 341 (2000) .

\bibitem{key-7}J. P. Keating, Nonlinearity \textbf{4}, 277 (1991).

\bibitem{hanay} J.H. Hannay and M.V. Berry, Physica D \textbf{1},
267 (1980).

\bibitem{opetor} A.M.F. Rivas and A.M. Ozorio de Almeida, Annals
of Physics \textbf{276,} 223 (1999).

\bibitem{matos} M. Matos and A.M. Ozorio de Almeida, Annals of Physics
\textbf{237}, 46 (1995).

\bibitem{Eckhardt} B. Eckhardt, J. Phys. A: Math. Gen. \textbf{19},
1823 (1986).

\bibitem{nonen} S. Nonnenmacher, Nonlinearity \textbf{10}, 1569 (1997).

\bibitem{scarelem}A.M.F. Rivas, J. Phys. A: Math. Gen{\tiny . }\textbf{46},
145101 (2013).

\bibitem{6ver}E. G. Vergini, J. Phys. A: Math. Gen., \textbf{33},
4709 (2000); E. G. Vergini and G. G. Carlo, J Phys. A: Math. Gen.,
\textbf{33,} 4717 (2000).

\bibitem{hetero} D.A. Wisniacki, E.G. Vergini, R.M. Benito and F.
Borondo, Phys. Rev. Lett.\textbf{ 94}, 054101 (2005).

\bibitem{13ver} E. G. Vergini, D. Schneider and A. M. F. Rivas, J.
Phys. A: Math. Theor.\textbf{ 41,} 405102 (2008).

\bibitem{ribeiro} A.D. Ribeiro, M.A.M. de Aguiar, and M. Baranger,
Phys. Rev. E.\textbf{ 69}, 066204 (2004).\end{thebibliography}
\end{document}